\documentclass[graybox]{svmult}

\usepackage{type1cm}        
\usepackage{makeidx}         
\usepackage{graphicx}        
\usepackage{multicol}        
\usepackage[bottom]{footmisc}

\usepackage{newtxtext}       %
\usepackage{newtxmath}       

\makeindex             

\usepackage[utf8]{inputenc}
\usepackage[english]{babel}
\usepackage{amsmath,graphicx,enumerate,empheq}
\usepackage{epstopdf}
\usepackage{multirow}
\usepackage{hyperref}
\usepackage{color}
\usepackage{bm}
\usepackage[a4paper,includeheadfoot,margin=2.54cm]{geometry}
\usepackage{mathtools}
\usepackage{authblk}

\begin{document}

\title*{The many faces of fluctuation-dissipation relations\\ out of equilibrium}

\author{Marco Baldovin and Lorenzo Caprini and Andrea Puglisi and Alessandro Sarracino and Angelo Vulpiani}
\institute{
Marco Baldovin \at Dipartimento di Fisica, Universit\`a di Roma  Sapienza, P.le Aldo Moro 2, 00185, Rome, Italy, \email{baldovin.m@gmail.com}
\and 
Lorenzo Caprini \at Scuola di Scienze e Tecnologie, Università di Camerino, Via Madonna delle Carceri, I-62032, Camerino, Italy, \email{lorenzo.caprini@gssi.it}
\and
Andrea Puglisi \at Istituto dei Sistemi Complessi - CNR and Dipartimento di Fisica, Universit\`a di Roma Sapienza, P.le Aldo Moro 2, 00185, Rome, Italy, \email{andrea.puglisi@cnr.it}
\and
Alessandro Sarracino \at Dipartimento di Ingegneria, Universit\`a della Campania ``L. Vanvitelli'', via Roma 29, 81031 Aversa (Caserta), Italy, \email{alessandro.sarracino@unicampania.it}
\and
Angelo Vulpiani \at  Dipartimento di Fisica, Universit\`a di Roma  Sapienza, P.le Aldo Moro 2, 00185, Rome, Italy, \email{angelo.vulpiani@roma1.infn.it}
}

\maketitle

\abstract{
  In this paper we offer to the reader an essential review of the
  theory of Fluctuation-Dissipation Relations (FDR), from the first
  formulations due to Einstein and Onsager, to the recent developments
  in the framework of stochastic thermodynamics of non-equilibrium
  system. We focus on two general approaches, somehow complementary,
  where out-of-equilibrium contributions to the FDR are expressed in
  terms of different quantities, related either to the stationary
  distribution or to the transition rates of the system. In
  particular, we discuss applications of the FDR in the general field
  of causation and inference, and in the contexts of non-equilibrium
  systems, such as spin models, granular media and active matter.
}



\vspace{1cm}
\section{Introduction}
\label{intro}

The Fluctuation-Dissipation Relation (FDR) is among the few pillars of
non-equilibrium statistical mechanics.  The reason of its great
relevance is rather transparent: it allows to compute the statistical
response of a system to small external perturbations in terms of
correlations of the unperturbed dynamics.  In other words, one can
understand how the system reacts to an external disturbance just
looking at the statistical features in the absence of any perturbation: in
such a way it is possible to determine perturbed properties
(response) in terms of unperturbed features (correlations).

The FDR has been widely investigated in the context of turbulence (and
more generally statistical fluid mechanics): for instance it plays a
key role for the closure problem in the Kraichnan's
approach~\cite{Kr59}. Moreover, there is a wide interest of the scientific
community active in geophysical systems, in particular, for climate
dynamics, where it is very important to understand the features of the
system under perturbations (such as a volcanic eruption, or a change
of CO$_2$ concentration) in terms of the knowledge based on time
series.  Another very relevant field where the FDR has been used and
investigated is the general theory of stochastic thermodynamics, with
particular focus on models for colloidal systems, granular media and
active matter. Finally, FDRs play a central role in the study of the non-equilibrium dynamics of slow relaxing systems, such as Ising models or spin glasses.
 

Since response and dissipation are intimately related (this intuitive
fact is made more formal later in this section), in this paper we use
``Fluctuation-Response'' and ``Fluctuation-Dissipation'' in an
interchangeable way.  Historically, one of the first examples of
Fluctuation-Response relation is given by the formula expressing the
fluctuations of energy in an equilibrium system at temperature $T$
with a (constant volume) heat capacity $C_v$, that reads
\begin{equation} \label{enflu}
\langle E^2 \rangle-\langle E\rangle^2=k_B T^2 C_v.
\end{equation}
On the left hand side of the formula one has the fluctuations in an
unperturbed system, while on the right hand side there is a quantity
representing a response (the heat capacity), and the factor of
proportionality between the two is the temperature ($k_B$ is the
Boltzmann constant). Einstein derived an analogous formula connecting
the diffusivity $D$ to the mobility $\mu$ for a Brownian particle
dispersed in a solvent fluid at thermodynamic equilibrium:
\begin{equation} \label{einst}
D= k_B T \mu,
\end{equation}
where again the unperturbed fluctuations (diffusivity) are
proportional to response (mobility) through a factor of
proportionality represented by the bath temperature $T$.

The two previous examples are instances of the larger class of
so-called ``static'' equilibrium FDR, as they do not involve
time-dependent quantities. In the first half of the $20$th century a
series of experimental and theoretical works made longer and longer
the list of such kind of relations, always tying in the same way
spontaneous fluctuations, response and
temperature~\cite{KTH91,BPRV08}. A noticeable example from this list
is the expression given by Nyquist in 1938, relating the fluctuations
of voltage in a conducting wire where no potential differences or
currents are externally applied (the so-called Johnson noise) to the
resistance of the conductor and the temperature. The resistance is the
analogous of the mobility and of the heat capacity in the previous
equations, i.e. it represents a response. In this case it is also
particularly simple to appreciate the equivalence between response and
dissipation.

A first step towards the generalisation to a time-dependent - or
dynamic - relation is represented by the regression hypothesis made by
Onsager in 1931~\cite{O31,O31b}, which states that - for small
perturbations from equilibrium - the system returns to equilibrium at
the same rate as a fluctuation does at equilibrium.  This fact is
already contained in the Einstein relation above. By recalling the
general connection between diffusivity and the velocity autocorrelation,
i.e. that
\begin{equation}
D=\int_0^{\infty} dt \langle v(t)v(0)\rangle,
\end{equation}
we can transform  Eq.~\eqref{einst} into
\begin{equation} \label{einst2}
\langle v(t)v(0)\rangle= k_B T  R_{v F}(t),
\end{equation}
with the identification
\begin{equation} \label{mob}
\mu = \int_0^\infty dt R_{vF}(t).
\end{equation}
In the r.h.s. of Eq.~\eqref{mob} we define the so-called response
function, $R_{vF}(t)$, which connects the mean variation of the
particle's velocity at time $t$ with a perturbation of the external
force applied at time $0$.

In order to discuss in full generality the FDR, we need to give a
general definition of response function, which is the central object
of linear response theory. We restrict the discussion to the linear
perturbation of stationary states, i.e. states which are invariant
under translations of time, so that time-dependent correlation
functions and response functions only depend on differences of
times. Generalisations to non-steady states are mentioned in Section~\ref{2fdr}.

The response function $R_{\mathcal{O}\mathcal{F}}(t)$ of the
observable $\mathcal{O}(t)$ to a time-dependent perturbation of a
parameter or degree of freedom $\delta \mathcal{F}(t)$ is implicitly
defined in the following relation
\begin{equation}
\overline{\Delta \mathcal{O}(t)} = \int_{-\infty}^t dt' R_{\mathcal{O}\mathcal{F}}(t-t') \delta\mathcal{F}(t'),
\end{equation}
where $\overline{\Delta \mathcal{O}(t)} =
\overline{\mathcal{O}(t)}-\langle\mathcal{O}(t) \rangle_0$ represents
the average deviation, at time $t$, of the observable $\mathcal{O}$
with respect to its average value in the unperturbed stationary
system. Here $\overline{f(t)}$ denotes an average of the observable $f$
at time $t$ over many realisations of the same perturbation, while
$\langle f \rangle_0$ denotes the average of $f$ in the stationary
unperturbed state, which is not time-dependent. It is clear that,
taking an impulsive shape for the external perturbation,
i.e. $\delta\mathcal{F}(t)=\Delta \mathcal{F}\delta(t)$ (with
$\delta(t)$ the Dirac delta distribution), one has
\begin{equation}
\frac{\overline{\Delta \mathcal{O}(t)}|_{imp}}{\Delta \mathcal{F}} = R_{\mathcal{O}\mathcal{F}}(t),
\label{imp}
\end{equation}
which is also an operational definition of the response function. Here
we stress that $\Delta \mathcal{F}$ has the dimensions of a
time-integral of $\mathcal{F}(t)$. When (for instance) the
perturbation has the shape of a Heaviside unit step function,
i.e. $\delta\mathcal{F}(t)=\delta \mathcal{F}_0 \Theta(t)$, then
\begin{equation}
\frac{\overline{\Delta \mathcal{O}(t)}|_{step}}{\delta \mathcal{F}_0} = \int_0^t dt' R_{\mathcal{O}\mathcal{F}}(t').
\end{equation}
If $\mathcal{O}(t)$ is the tracer's velocity along one axis and
$\mathcal{F}(t)$ is the external force applied from time $0$ to time
$\infty$ to the tracer (parallel to that axis), the final velocity
reached by the tracer is exactly $\delta \mathcal{F}_0\int_0^\infty
dt' R_{vF}(t')$, which explains the connection with the identification made
in Eq.~\eqref{mob}.

The FDR for systems with  Hamiltonian $\mathcal{H}$ at equilibrium  with a thermostat at temperature
$T$ -- historically attributed to Callen and Welton and immediately after generalised by Kubo ~\cite{K57} -- reads:
\begin{equation} \label{fdt}
R_{\mathcal{O}\mathcal{F}}(t) = \frac{1}{k_B T}\langle \mathcal{O}(t) \dot{A}(0)\rangle_0 = -\frac{1}{k_B T}\langle \dot{\mathcal{O}}(t) A(0)\rangle_0,
\end{equation}
where $A$ is the observable (or degree of freedom) which is coupled to
$\mathcal{F}(t)$ in the Hamiltonian to produce the perturbation,
i.e. $\mathcal{H}(t)=\mathcal{H}_0 - \mathcal{F}(t)A$. It is easy to
verify that if $\mathcal{O}$ is the tracer's velocity and
$\mathcal{F}(t)$ is an external force applied to its $x$ coordinate,
Eq.~\eqref{fdt} becomes Eq.~\eqref{einst2}. In conclusion the
``dynamical" Einstein relation is a particular case of equilibrium
FDR. From Eq.~\eqref{fdt} one may get several possible variants, which
are useful in different physical situations. A large amount of
remarkable results concern, for instance, the time-Fourier transform
of Eq.~\eqref{fdt}, as well as the relation connecting currents/flows
and transport coefficients in spatially extended systems (the
so-called Green-Kubo relations, see below)~\cite{KTH91,BPRV08}.

The equilibrium FDR is valid also in the framework of stochastic
processes, when they describe the dynamics of system fluctuating
around thermal equilibrium. The main difference with the case
considered by Kubo is that a stochastic process typically describes
small systems, far from the thermodynamic limit, but the system size
is in fact irrelevant for the purpose of the validity of the FDR. In
the case of large systems (without long-range correlations), however,
the averages are easily taken by means of one or few experiments,
while in a stochastic process where noise is large, one needs to average
over many realisations.  An illustrative example is the so-called
Klein-Kramers model which describes the dynamics of simple particle
systems at thermal equilibrium~\cite{G90}. In one dimension its
stochastic differential equations read:
\begin{subequations} \label{KK}
\begin{align} 
\frac{d x(t)}{dt} &= v(t) \\
m\frac{d v(t)}{dt} &= -\frac{d U(x)}{dx} -\gamma v(t) + \sqrt{2 \gamma k_B T}\xi(t),
\end{align}
\end{subequations}
where $\xi(t)$ is a white Gaussian noise with $\langle \xi(t) \rangle
=0$ and $\langle \xi(t)\xi(t') \rangle = \delta(t-t')$, $\gamma$ is
the viscous damping, $U(x)$ is an external potential. The model can be
easily generalised to $N>1$ interacting particles in any
dimensions. In the absence of the external potential, Eq.~\eqref{KK}
coincides with the original Langevin equation proposed a few years
after the theories of Einstein~\cite{E05} and Smoluchovski~\cite{s06}
to explain diffusion in Brownian motion~\cite{L08}. Its steady
probability distribution (achieved with the condition $\gamma >0$ and
confining potential) is given by $P(x,v) \propto
e^{-\mathcal{H}(x,v)/(k_B T)}$ with $\mathcal{H}(x,v)=mv^2/2 +
U(x)$. Linear response theory, when applied to the Klein-Kramers model
in its stationary state, gives exactly the same result as
Eq.~\eqref{fdt}~\cite{R89,BPRV08}. The Klein-Kramers process is
Markovian with respect to the variables $(x,v)$, a property which is a
rough approximation for the dynamics of a tracer which interacts with
other particles in a fluid. Typically it has to be generalised to take
into account retarded (hydrodynamic) effects, by the introduction of
linear memory terms, e.g. by writing a Generalized Langevin Equation
(GLE)~\cite{KTH91}:
\begin{equation} \label{gle}
m \frac{d v(t)}{dt} = -\int_{-\infty}^t dt' \Gamma(t-t') v(t') + \eta(t),
\end{equation}
where $\Gamma(t)$ is a memory kernel representing retarded damping,
and $\eta(t)$ is a stationary stochastic process with zero average
$\langle \eta(t) \rangle =0$. The noise time-correlation -- to comply
with the requirement of thermodynamic equilibrium (i.e. steady Gibbs
distribution and detailed balance) -- must satisfy the so--called FDR of
the second kind:
\begin{equation} \label{fdt2}
\Gamma(t) = \frac{1}{k_B T} \langle \eta(t) \eta(0) \rangle.
\end{equation}
It is clear that Eq.~\eqref{fdt2} has the same structure of
Eq.~\eqref{fdt} and this motivates the name of the relation. The
Markovian case (damping with zero memory) is obtained when
$\Gamma(t)=2\gamma \delta(t)$ (recalling that $\int_{0}^t dt'
2\gamma \delta(t') v(t')=\gamma v(t)$).  For a more detailed
discussion of the significance of this condition and its connection to
detailed balance, we invite to read Section~4.1 of~\cite{temprev}.

This brief review paper is organised as follows. In Section~\ref{2fdr}
we introduce two different possible approaches to the FDR, which are
based either on the knowledge of the stationary distribution or on the
knowledge of the dynamical rules of the model. Then, in
Section~\ref{appl}, we discuss several applications of the FDR, in
particular in the field of non-equilibrium systems, such as granular
media and active matter. Finally, in Section~\ref{concl}, some
conclusions are drawn.

\section{Two  approaches to non-equilibrium FDR}
\label{2fdr}

The first examples of FDR date back to Einstein’s work on Brownian
motion (1905), and to Onsager's regression hypothesis (1930's).  Since
initially the FDR was obtained for Hamiltonian systems in
thermodynamic equilibrium, somehow there is a certain confusion on its
real validity.  Here we summarise two different generalizations of FDR
which both hold for a broad class of systems, including the non
equilibrium cases~\cite{BPRV08}.

\subsection{An approach based upon the knowledge of the stationary distribution}

Let us consider a system whose stationary probability density
$P_{st}({\bf x})$ is non-vanishing everywhere, and wonder about the
time behavior of the mean response of the variable $x_n(t)$ at time
$t$ under a small impulsive perturbation $\delta {\bf x}(0)$.  We can
write
 $$
 \overline{\delta x_n(t)}= \Big\langle x_n(t) \Big\rangle_p - \Big\langle x_n(t) \Big\rangle 
 $$
 where $\Big\langle \,\, \Big\rangle_p$ and $ \Big\langle \,\,
 \Big\rangle$ denote the average for the perturbed and the unperturbed
 systems respectively.  For a Markov system we can write
$$
\Big\langle x_n(t)  \Big\rangle_p =\int  x_n P_p({\bf y}) W({\bf y} \to {\bf x},t) \, d{\bf x}d{\bf y} \,\,\, ,\,\,\,
 \Big\langle x_n(t)  \Big\rangle =\int  x_n P_0({\bf y}) W({\bf y} \to {\bf x},t) \, d{\bf x}d{\bf y} \,\,\, ,\,\,\,
$$
 where $W({\bf y} \to {\bf x},t)$ is the  probability  of a transition from  ${\bf y}$ at time  $0$ to
 ${\bf x}$ at time  $t$,
  $P_0({\bf y})=P_{st}({\bf y})$ and  $P_p({\bf y})$ is the initial distribution of the perturbed system.
 
In the case of an impulsive perturbation, the perturbed probability satisfies  $P_p({\bf y})=P_{st}({\bf y} -\delta {\bf x}(0))$ which allow us to derive a compact expression for $\overline{\delta x_n}$ when the perturbation is small:
\begin{equation} \label{LN}
\overline{\delta x_n(t)}= - \sum_j \Big\langle x_n(t) { \partial \ln P_{st}[{\bf x}(0)]  \over \partial x_j(0)}   \delta x_j(0)  \Big\rangle\,,
\end{equation}
where the average is performed in the unperturbed system.  Let us note
that the assumption of small perturbation is necessary only in the
last step of the derivation of Eq.~\eqref{LN} therefore, such a result
can be generalized to the case of non-infinitesimal $\delta {\bf
  x}(0)$~\cite{BLMV03}. As by-product we have that it is possible to
avoid the criticism of van Kampen according to which it is not
possible to rely on an expansion for small perturbations, because
chaos makes them grow exponentially~\cite{van1971discussion}.  On the
contrary, in the derivation of the above
result~\cite{falcioni1990correlation}, there are only assumptions
about $\delta {\bf x}(t=0)$ and therefore chaos has no relevance.

We can say that formula \eqref{LN} summarizes the main results of the
linear theory, e.g. in Hamiltonian systems and stochastic processes:
in addition one understands the existence of a link between response
and a suitable correlation function even in non-equilibrium
systems~\cite{BPRV08}.  For instance in inviscid hydrodynamics with an
ultraviolet cutoff, in spite of the non trivial dynamics, since the
presence of quadratic invariant, and a Liouville theorem, one has a
Gaussian statistics and therefore a FDR holds for each of the
variables, i.e. self-response functions to infinitesimal perturbations
coincide with the corresponding self-correlation functions.  Let us
note that although $P_{st}({\bf
  x})$ is Gaussian the dynamics is non
linear and it is not easy to compute the correlation functions.

Beyond the many conceptual advantages of eq. \eqref{LN} there is an
obvious practical limit: the difficulty to determine $P_{st}({\bf
  x})$, which is known only in some specific cases.  In the next
subsection we will discuss an approach which does not need the knowledge
of $P_{st}({\bf x})$.

Let us open a brief parenthesis on chaotic deterministic dissipative
systems: because of the phase space contraction one has that the invariant
measure is singular, typically with a multifractal structure,
and therefore, Eq. \eqref{LN} cannot be applied.  A quite natural
temptation is to add a small amount of noise, so that a smoothing of
the invariant probability density allows for the use of the FDR.  At a
first glance such an approach can appear unfair. On the contrary the
idea of the beneficial role of the noise, which seems to date back to
Kolmogorov, has strong bases: a small noisy term in the evolution
equations has the role of selecting the natural measure: one can say
that in the numerical experiments the round-off errors of the computer
play a positive role.  It is quite natural to expect that the behaviors
of the purely deterministic chaotic system are very close to those obtained by adding a small amount of noise; such a conjecture is widely
confirmed by numerical computations~\cite{er}.

A similar approach was extended by Seifert and Speck, who established
interesting connections of the FDR with observables in the framework
of stochastic thermodynamics, such as entropy production and
housekeeping heat~\cite{ss06,seifert2010fluctuation,seifertrev} (see also the next Section).

\subsection{An approach based upon the knowledge of the dynamical model}

\label{SubSec:CapriniStatement}

When the dynamics of the system under study is defined in terms of
transition rates or Langevin equations, but the stationary probability
density function is not known, a complementary approach with 
respect to the one discussed in the previous subsection can provide a FDR
valid also out of equilibrium.  These kinds of FDRs have been derived in
several different contexts, following different mathematical schemes
(see discussion below). 

The general approach dates back to the 60's of the 20th century, when
Furutsu and Novikov independently derived, under general conditions, a
FDR~\cite{n65,furutsu1964statistical} which expresses the response
function of a Gaussian process in terms of the equilibrium
time-correlation between the observed variable and the Gaussian noise
itself.  Nowadays, a method based on similar ideas - sometimes termed
Malliavin weight sampling~\cite{malliavin} - has been extended to
include field theories through the Martin-Siggia-Rose-Jansen-de
Dominicis approach~\cite{cugliandolo1996large,
  andreanov2006dynamical,aron} and employed in the context of
particle-based glassy systems to numerically calculate effective
temperature and
susceptibility~\cite{cugli1,cuglirev,crisanti2003violation}. This
allows one to express the response function in terms of suitable
correlation functions of the state variables.  We mention here
examples for non-equilibrium Langevin dynamics driven by a
time-dependent force both in
overdamped~\cite{seifert2010fluctuation,baiesi2009fluctuations,baiesi2009nonequilibrium,BMW10,yolcu2017general}
and underdamped regimes~\cite{baiesi2010nonequilibrium}, or even in
the presence of a non-linear Stokes
force~\cite{sarracino2013time}. The non-equilibrium terms appearing in
the generalized FDR have been interpreted in several ways: some authors focused on the different roles of entropic and frenetic
contributions (for a recent review, see~\cite{maes}), outlining their
different nature with respect to the symmetry under time-reversal
transformation; other approaches have focused on the connection 
with entropy production and heat~\cite{harada1,ss06,LBS14}.

The class of generalized FDR so far mentioned is expressed in terms of
the correlation between the observable $\mathcal{O}$ and a function of
both the state variables and their time-derivatives. Without loss of
generality, the starting point for these relations is of the form:
\begin{equation}
\label{eq:Novikovequation}
R_{\mathcal{O}x_j}(t,s)= \langle \mathcal{O}(t) \mathcal{M}_j[\mathbf{x}(s), \dot{\mathbf{x}}(s)]  \rangle \,,
\end{equation} 
where, as usual, the average in the r.h.s. of
Eq.~\eqref{eq:Novikovequation} is performed through the unperturbed
measure.  $\mathcal{M}_j$ is a function uniquely determined by the
dynamics of the system under consideration that depends both on
$\mathbf{x}$ and $\dot{\mathbf{x}}$. Its functional form can be
expressed in terms of known observables: for instance, in the case of
continuous first order dynamics of the kind
\begin{equation}
\dot x_j =
f_j(\mathbf{x})+\sqrt{2D_j}\eta_j,
\end{equation}
where $\eta_j$ is a white noise with zero
average and unit variance, one has
\begin{equation}
  \mathcal{M}_j=\frac{1}{2D_j}(\dot{x}_j - f_j).
  \end{equation}
 The
above result is general, holding in stationary or transient
non-equilibrium regimes.  In some cases, i.e. when the quantity
$\mathcal{M}_j$ can be measured, Eq.~\eqref{eq:Novikovequation} may
represent an advantage with respect to Eq.~\eqref{LN} (which depends
upon the knowledge of the steady-state probability). 

The explicit dependence on the time-derivative of the state variables,
$\dot{\mathbf{x}}$, in Eq.~\eqref{eq:Novikovequation} may still
represent a source of complications.  Restricting to the calculation
of the response matrix, $R_{x_i x_j}(t)$, i.e. such that
$\mathcal{O}=x_i$, from Eq.~\eqref{eq:Novikovequation} one can
derive~\cite{caprini2021generalized} a simpler expression for
processes with additive Gaussian noises in the stationary state (the
result can be easily generalized to the case of non-diagonal
diffusion, not reported for conciseness):
\begin{equation}
\label{eq:R_capriniadditive}
R_{x_ix_j}(t)= - \frac{1}{2D_j}\left[ \langle x_i(t) f_j(0)  \rangle + \langle f_i(t) x_j(0)  \rangle \right]\,.
\end{equation}
Each element of the response matrix is given by the sum of two
correlations: i) the temporal correlation between the observed
variable and the force ruling the dynamics of the perturbed variable
and ii) the temporal correlation between the force of the observed
variable and the perturbed variable (that for the diagonal elements,
$R_{x_ix_i}(t)$, is the same correlation of i) with swapped times).
The two terms are equal only at equilibrium. On the contrary they
differ when detailed balance does not hold.  Note that the generalized
FDR~\eqref{eq:R_capriniadditive}, differently from the
forms~\eqref{LN} or~\eqref{eq:Novikovequation}, is not determined by
the time-correlation between the observed variable evaluated at $t$
and another observable at $s<t$.  Moreover, path-integral FDRs require
the explicit knowledge of the microscopic dynamics, at variance
with the approach~\eqref{LN} which only requires a model of the
stationary probability in phase space: it must be noticed that in
experimental situations it can be simpler to formulate a model for the
steady state probability rather than for the full dynamics. In both
cases, however, one needs to individuate the relevant variables, an
often underestimated aspect~\cite{GPSV14}.

The generalized FDR~\eqref{eq:R_capriniadditive} is particularly
fascinating because the diagonal elements of the response matrix
(r.h.s of Eq.~\eqref{eq:R_capriniadditive}) are expressed in terms of
the time-symmetric part of the anticipated/retarded equipartition
relations while the non-diagonal elements represent the time-symmetric
part of the anticipated/retarded Virial
equation~\cite{caprini2021generalized}.  Indeed, because of the
causality condition, we have $R_{x_i x_j}(t=0)=\delta_{ij}$, so that
the initial time elements of the response matrix contain the same
information as the generalized equipartition and Virial equations
holding out of equilibrium, namely:
\begin{flalign}
D_i= \langle x_i f_i\rangle \,, \,\,\;\;
\langle x_i f_j\rangle=-\langle x_j f_i\rangle \,.  
\end{flalign}
This physical interpretation has been discussed in detail
in~\cite{caprini2021generalized} and exploited in well-known examples,
such as passive and active colloids both in underdamped and overdamped
regimes, see also Section~\ref{applact}.



Let us also comment on the interesting case of discrete variables,
relevant for instance for the Ising model or spin glasses, which
requires some care. In particular, for spins $\sigma_i=\pm 1$, with
$i=1,\ldots,N$, evolving according to a Master Equation with
unperturbed transition rates form the configuration $\sigma$ to the
configuration $\sigma'$, $w(\sigma\to\sigma')$, in contact with a
reservoir at temperature $T$, the response of an observable $
\mathcal{O}(\sigma)$ at a magnetic field $\mathcal{F}$ switched on at
time $s$ on site $j$ takes the following form~\cite{LCZ05}
\begin{equation}
R_{\mathcal{O}\mathcal{F}}(t,s)=\frac{1}{2T}\left\{\frac{\partial}{\partial
  s}\langle \mathcal{O}(t)\sigma_j(s)\rangle-\left\langle
\mathcal{O}(t)B_j(s)\right\rangle\right\},
\label{discreto}
\end{equation}
where the quantity $B_j[\sigma]$ is defined by
\begin{equation}
B_j[\sigma(s)]=\sum_{\sigma'}[\sigma_j'-\sigma_j(s)]w[\sigma(s)\to\sigma'].
\label{B}
\end{equation}
The equilibrium FDT~(\ref{fdt}) is obtained exploiting the property
\begin{equation}
\left\langle
\mathcal{O}(t)\sum_{\sigma'}[\sigma_j'-\sigma_j(s)]w[\sigma(s)\to\sigma']\right\rangle_{eq}=-\frac{\partial}{\partial
  s}\langle \mathcal{O}(t)\sigma_j(s)\rangle_{eq},
\label{eqprop}
\end{equation}
valid when the average is taken at equilibrium~\cite{LCZ05,LCSZ08b}.

\section{Applications} 
\label{appl}

In this Section we discuss recent applications of the generalised
formulae discussed above to different problems. We start with two more
theoretical cases, namely the broad class of spin and disordered
systems and the search for causality measurements, and we conclude
with applications to paradigmatic macroscopic physical systems, that
are granular and active systems, where the dynamics of each particle
is intrinsically out of equilibrium.

\subsection{The interesting case of causation through response}
\label{bald}
\DeclareRobustCommand{\vect}[1]{
  \ifcat#1\relax
    \boldsymbol{#1}
  \else
    \mathbf{#1}
  \fi}
  
Among the many practical applications of the generalized
FDR~\eqref{LN}, its use in the field of causal inference has a
particular conceptual interest. It is well known that, in order to
understand the cause-effect relations holding between different
elements of a system, measuring the degree of correlation of the
variables may be, in general, of little help: two elements can be
highly correlated even in the absence of a causal link, as summarized
by the notorious adage ``correlation does not imply causation''. The
right way to characterize causal relations is indeed to \textit{probe}
the system under study, i.e. to perturb it in some way and to observe
the effects of this external action, comparing them to the usual
behavior of the system in the absence of
perturbation~\cite{barnett09,Aurell2016causal}; this is, for instance,
the fundamental idea at the basis of Pearls' formalism of
counterfactual inference~\cite{PearlBook}. When dealing with physical
systems, as discussed in the Introduction, the effect of an external
perturbation is quantified by response functions, which are therefore
natural indicators of causal
relations~\cite{Aurell2016causal,baldovin2020}. In this respect, a
surprising consequence of Eq.~\eqref{LN} is that these observables can
be estimated by measuring proper correlation functions in an
\textit{unperturbed} dynamics: in other words, the generalized FDR
allows to infer causal relations without operating any external action
on the system, i.e. without actually probing it.

\begin{figure}
 \centering
 \includegraphics[width=.7\linewidth]{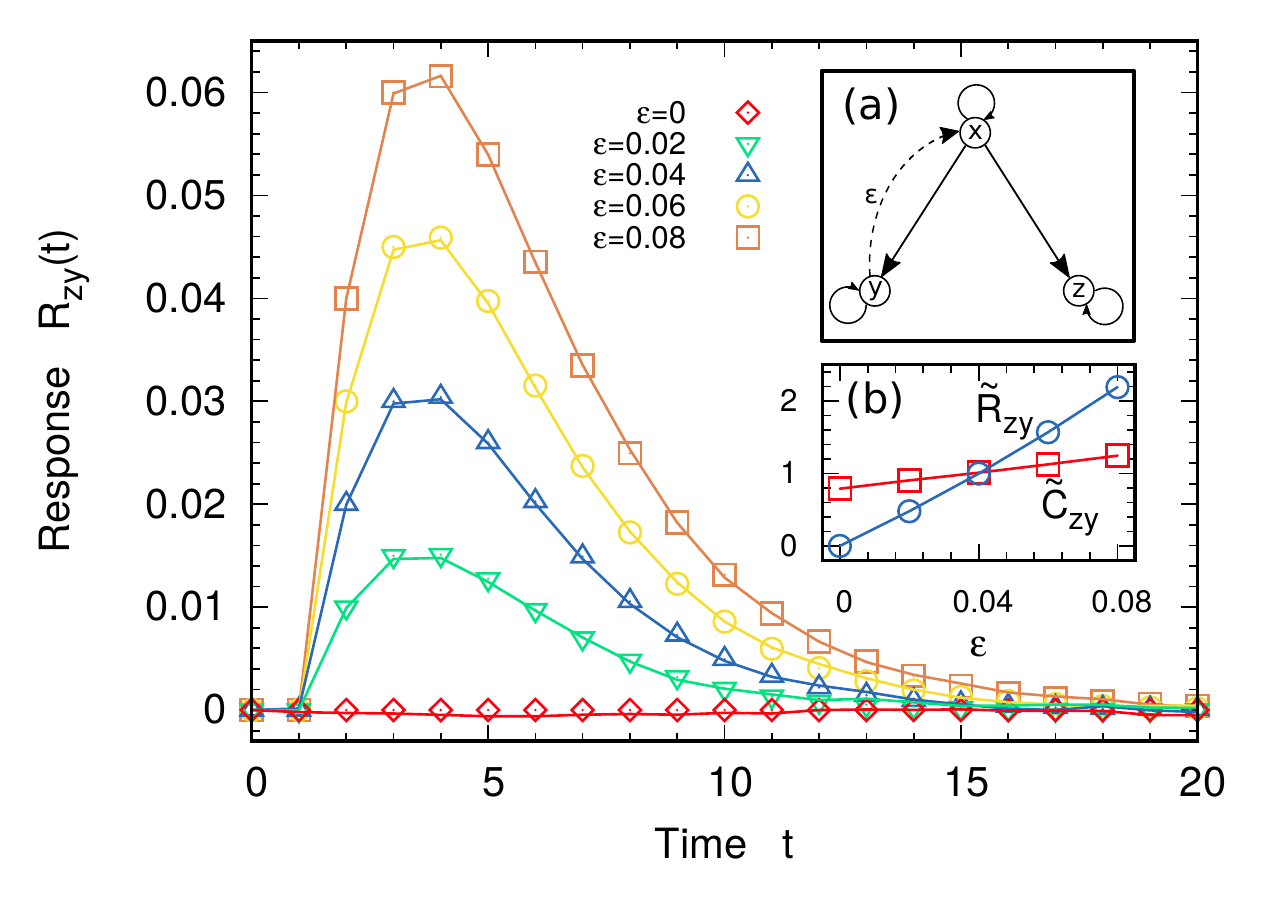}
 \caption{\label{fig:caus} Relation between causation and response. Main plot: response matrix element $R_{zy}(t)$ of model~\eqref{eq:chocolate}, as a function of time, for several values of the feedback parameter $\varepsilon$; numerical simulations in which the system is actually perturbed (points) are compared to the predictions of the generalized FDR~\eqref{LN} (lines). Inset (a): scheme of the interactions occurring in model~\eqref{eq:chocolate}. Inset (b): correlations (red squares) and responses (blue circles) integrated over time, as functions of $\varepsilon$; both quantities are rescaled with their values at $\varepsilon=0.04$ for graphical convenience. Parameters: $a=0.5$, $b=1$. Perturbation for the computation of response: $\delta y_0=0.01$. $M=10^6$ trajectories have been considered for the averages.}
\end{figure}
 
To show the above point, let us consider the example of a linear stochastic 
dynamics for the three-dimensional vector $(x_t, y_t, z_t)$ in discrete time, 
ruled by the following Markov process:
\begin{subequations} 
\begin{align}
  \label{eq:chocx} x_{t+1}=&a x_t + \varepsilon y_t  + b \eta^{(x)}_t \\
  \label{eq:chocy} y_{t+1}=&a x_t + a y_t + b \eta^{(y)}_t \\
  \label{eq:chocz} z_{t+1}=&a x_t + a z_t + b \eta^{(z)}_t
\end{align}
\label{eq:chocolate}
\end{subequations}
where $a$, $b$ and $\varepsilon$ are suitable constants and $\eta^{(x)}_t$, 
$\eta^{(y)}_t$, $\eta^{(z)}_t$ are independent, delta-correlated Gaussian 
variables with zero mean and unitary variance. In this model the dynamics of 
$y_t$ and $z_t$ is influenced by $x_t$, which feels in turn the effect of $y_t$ 
because of the feedback term $\varepsilon y_t$ in the r.h.s. of 
Eq.~\eqref{eq:chocx}. A sketch of the interaction scheme is shown in the inset 
(a) of Fig.~\ref{fig:caus}.

The main plot in Fig.~\ref{fig:caus} shows the time dependence of the response 
function between $y_t$ and $z_t$. As it is clear from the structure of the 
dynamics, after one time step there is no causal influence (an external 
perturbation of $y_t$ does not reflect on $z_{t+1}$). At subsequent times the 
dynamics of $z_t$ is altered by the perturbation, and the value of the response 
function crucially depends on the feedback parameter $\varepsilon$, as expected. 
Due to the linearity of the model, Eq.~\eqref{LN} can be simplified 
into~\cite{BPRV08}:
\begin{equation}
\label{eq:respcorr}
 R_t=C_t C^{-1}_0
\end{equation}
where $C_t$ represents the correlation matrix at time $t$, i.e. $C_t^{ij}=\langle x_i(t) x_j(0)\rangle$ (with $x_1=x,x_2=y,x_3=z$), and $C^{-1}_0$ is the 
inverse of $C_0$. The linearity of Eqs.~\eqref{eq:chocolate} implies that $P_{st}$ is a multi-variate Gaussian and this immediately leads to Eq.~\eqref{eq:respcorr}. Exploiting this version of the generalized FDR, as shown in 
Fig.~\ref{fig:caus}, one can estimate $R_{zy}(t)$ from a suitable combination of 
correlation functions: the agreement with the actual responses, computed from
numerical simulations, is excellent.

It is worth noticing that the mere knowledge of $C(t)$ is not at all informative 
about the causal links among the elements of the system. For the considered 
model this fact can be qualitatively appreciated by looking at inset (b) of 
Fig.~\ref{fig:caus}, where we compare the behavior of 
$\tilde{R}_{zy}=\int_0^\infty R_{zy}(t)\, dt$ and 
$\tilde{C}_{zy}=\int_0^\infty C_{zy}(t)\, dt$ as functions of $\varepsilon$:
while the former quantity, in the 
considered $\varepsilon \ll 1$ regime, is almost proportional to $\varepsilon$, 
the latter does not crucially depend on the feedback parameter and it is 
different from zero also for $\varepsilon=0$. This difference is clearly due, in 
the considered example, to the common dependence of $y_t$ and $z_t$ on the 
variable $x_t$, inducing a ``spurious'' correlation between the two (meaning
that such a correlation does not unveil any causal link between the two 
processes).

Using the generalized FDR is not the only way to get some insight into the 
causal structure of a physical system without perturbing its dynamics. A widely 
employed method is due to Granger~\cite{Granger69} and relies on the computation of 
the forecasting uncertainty for a given variable of a system, using linear 
regression models; if it is possible to improve the prediction's precision
by including in the model a second, different variable of the system, one may assume a cause-effect relation 
between the two. A different approach (which has been shown to be equivalent to
Granger's method in the case of linear dynamics~\cite{barnett09,barrett10}) is based on the analysis of information transfer between the 
variables, a process quantified by the so-called \textit{transfer 
entropy} and by other related observables~\cite{shreiber00,bossomaier,runge12,sun15}.
Despite the useful information provided by these approaches,
response functions appear to be more accurate
in characterizing causal relations, at least from a physical point of view;
indeed they quantify the (average) consequence of an actual intervention
on the system, at variance with Granger's method and transfer entropy analysis, which
face the problem from the point of view of predictability and uncertainty~\cite{barrett13,barnett18, baldovin2020,manshour2021causality}. In this respect, generalized FDRs as Eq.~\eqref{LN}
are, to the best of our knowledge, the only way to deduce the causal structure of
a system, in a proper physical sense, by only observing its spontaneous evolution.

\subsection{Spin and disordered systems}


Here we focus on some applications of the FDR in the contexts of
spin models and disordered systems.  As already underlined, the main
aim of an FDR is to give a tool to calculate a response without
applying the perturbation. The direct calculation of a linear response
function, for instance in numerical simulations (but the same can be
true for experiments), is a very time-demanding task: indeed, the
signal fluctuations generally increase when the applied field is
small, a condition required for the linear regime to hold. Therefore,
the application of FDRs in numerical computations is an effective
shortcut to get information on the response function from the
measure of the correlations in the unperturbed state. This shortcut
has been frequently used to develop field-free algorithms in the context of spin
systems~\cite{CR03,chatelain,ricci,LCZ05,CLSZ10}, and
glasses~\cite{berthier} or active matter~\cite{szamel}.  Let us note
that, at variance with previous attempts, specifically designed for a
numerical implementations~\cite{chatelain,ricci,berthier}, the FDR
reported in Eq.~(\ref{discreto}) involves the quantity $B$ defined
in~(\ref{B}), which is an observable quantity because only depends on
the state of the system at a given time and therefore can be in
principle measured in real experiments.

\subsubsection{Non-linear FDRs}

The FDRs in the form (\ref{discreto}) can be also derived at nonlinear
orders in the perturbation, involving multi-point correlation
functions.  Non-linear response functions play a central role in the
context of glassy
systems~\cite{biroli1,biroli2,diezemann2012nonlinear,speck}, where
usually two-point correlators remain always short-ranged due to the
presence of disorder. In particular, in a spinglass the linear
susceptibility does not diverge at the critical temperature, whereas
non-linear susceptibilities show a divergence when the low temperature
phase is approached, signaling a growing amorphous order in the
system.  Therefore, the relation between nonlinear responses and
multi-point correlation functions can be an important tool in the
context, as initially proposed in~\cite{BB05}. A general derivation of
nonlinear FDRs valid for arbitrary order was presented
in~\cite{LCSZ08a,LCSZ08b}. We report here the form of the second-order
response for spin variables perturbed by two fields $\mathcal{F}_1$
and $\mathcal{F}_2$ at sites $j_1$ and $j_2$ at times $t_1$ and
$t_2$~\cite{LCSZ08b}
\begin{eqnarray}
R_{\mathcal{O}\mathcal{F}}^{(2)}(t,t_1,t_2)&\equiv&\left . \frac{\delta
  \langle \mathcal{O}(t)\rangle_\mathcal{F}}{\delta
  \mathcal{F}_1(t_1)\delta \mathcal{F}_2(t_2)}\right|_{h=0}\nonumber
\\ &=&\frac{1}{4T^2}\left\{\frac{\partial}{\partial
  t_1}\frac{\partial}{\partial t_2}\langle \mathcal{O}(t)
\sigma_{j_1}(t_1)\sigma_{j_2}(t_2)\rangle -\frac{\partial}{\partial t_1}\langle
\mathcal{O}(t) \sigma_{j_1}(t_1)B_{j_2}(t_2)\rangle \right.\nonumber \\ &-&\left
. \frac{\partial}{\partial t_2}\langle \mathcal{O}(t)
B_{j_1}(t_1)\sigma_{j_2}(t_2)\rangle+\langle \mathcal{O}(t)
B_{j_1}(t_1)B_{j_2}(t_2)\rangle\right\}.
\label{second}
\end{eqnarray}

Let us note that at equilibrium, exploiting the property~(\ref{eqprop}),
Eq.~(\ref{second}) simplifies to
\begin{eqnarray}
R_{\mathcal{O}\mathcal{F}}^{(2)}(t,t_1,t_2)=\frac{1}{2T^2}\Big\{\frac{\partial}{\partial
  t_1}\frac{\partial}{\partial t_2}\langle \mathcal{O}(t)
\sigma_{j_1}(t_1)\sigma_{j_2}(t_2)\rangle -\frac{\partial}{\partial t_2}\langle
\mathcal{O}(t) B_{j_1}(t_1)\sigma_{j_2}(t_2)\rangle\Big\},
\end{eqnarray}
with $t>t_1>t_2$. Therefore, the presence of the model-dependent
quantity $B$ is not canceled, making the higher order FDRs somehow
less general than the linear one. As suggested in~\cite{basu}
and~\cite{helden} this observation can provide information on the
dynamical rules governing the system from the study of the equilibrium
nonlinear responses.

Other interesting applications of nonlinear FDRs are related to the
study of the thermal response of the system (namely, a perturbation
applied to the noise intensity) as discussed in~\cite{FB16}, or in the
wide field of nonlinear optics and quantum
spectroscopy~\cite{mukamel1,mukamel2}.

\subsubsection{Effective temperature}

One of the main theoretical applications of the FDRs is the
possibility to introduce an effective temperature, from the ratio
between response and correlation. Review articles on this interesting
subjects are~\cite{leuzzi,cuglirev,temprev,zannetti,CLZ07}.  Here we
illustrate such a concept for a spin system, where the linear
susceptibility, using the FDR~(\ref{discreto}), can be written as
\begin{eqnarray}
\chi(t,t_w) \equiv \int_{t_w}^t ds R_{\sigma \mathcal{F}}(t,s) =
\frac{\beta}{2} \int_{t_w}^t ds \left [\frac{\partial}{\partial s} C(t,s)
-  \langle \sigma_i (t) B_i(s) \rangle \right ],
\label{eff1}
\end{eqnarray}
where $C(t,s)=\langle \sigma_i(t)\sigma_i(s)\rangle$ and $t_w$ is a
reference waiting time.  Observing that the quantity
\begin{equation}
\psi(t,t_w) =  \int_{t_w}^t ds \frac{\partial}{\partial s}C(t,s)  = 1- C(t,t_w),
\label{eff3}
\end{equation}
for fixed $t_w$, is a monotonously increasing function of
time, one can reparametrize $t$ in terms of $\psi$ and write $\chi(\psi,t_w)$.

In equilibrium, there is no dependence on the waiting time
$t_w$ and one obtains a linear parametric representation
\begin{equation} \chi(\psi) = \beta \psi,
\label{eff6}
\end{equation}
yielding
\begin{equation}
\beta = \frac{d\chi(\psi)}{d\psi}.
\label{eff7}
\end{equation}
Out of equilibrium, a nonlinear dependence can arise and
an effective temperature can be introduced generalizing Eq.~(\ref{eff7})
\begin{equation} 
\beta_{eff}(\psi,t_w) =\frac{ \partial \chi(\psi,t_w)}{\partial\psi},
\label{eff8}
\end{equation}
with $\beta_{eff} = 1/T_{eff}$.
Then one can define a Fluctuation-Dissipation ratio with respect to the temperature $T$ of the dynamics (after the quench)
\begin{equation} 
X(\psi,t_w)=\frac{T}{T_{eff}(\psi,t_w)}.
\label{eff9}
\end{equation}
which represents a measure of the deviation from equilibrium. In the
limit of large waiting time, the functional dependence of $X$ on the
correlation function can show different behaviors, shedding light on
the relevance of different characteristic time-scales in the system. A
detailed discussion of this quantity in the context of aging and
glassy systems can be found in~\cite{CR03}. More recent applications
of the FDR to equilibrium and non-equilibrium properties of spin
glasses have been reported in~\cite{Baity-Jesi1838}.

The concept of effective temperature has been also applied to systems
in the stationary state, such as driven granular media or active
particles (see for instance~\cite{mossa08,shakerpoor2021einstein}). In
this case, the problem is to understand the meaning and the role
played by the effective temperature. In some situations, usually when
the system is gently driven and the entropy production flux is small,
the relevant features of the system behavior can be successfully
interpreted in terms of this parameter, leading to an equilibrium-like
description. In other cases, the effective temperature can represent
an evocative or appealing concept, but does not significantly help in
the understanding of the underlying physical mechanisms, see next
Section.

\subsection{Granular materials}
\label{sub51}

Granular materials appear in our everyday life and in several
industrial applications, posing deep questions to statistical physics
and technology~\cite{JN92,andreotti13,puglio15,garzo2019granular}.  A granular medium is
an ensemble of macroscopic ``grains'', which interact (among each
other and with the surroundings) through non-conservative
forces. Several orders of magnitude separate the average energy of
internal thermal fluctuations at room temperature - $k_B T \sim 5
\cdot 10^{-21} J$ - and the macroscopic energy of a grain (e.g. that
related to the position and motion of center of mass): for instance $m
g r \sim 10^{-5} J$ for a steel sphere with $r=2 mm$, $g$ being the
gravity acceleration.  Granular media can display ``phase'' behaviors:
when diluted and under strong shaking a granular ``gas'' is realised,
but when allowed volume and/or the intensity of shaking are reduced,
the granular system behaves as a dense ``liquid'' or a slowly deforming
"solid"~\cite{JNB96b}. The slow-dense phase, close to the so-called
{\em jamming} transition, is difficult to be analysed:
we refer the reader to different theoretical
approaches~\cite{mehta89,edwards89c,edwards94,nowak98,bcl02,ono2002effective,makse04,RNDRB05,ciamarra06,baule2018edwards}. We
briefly summarise the more clear situation established for granular
gases and liquids.

A granular gas is realized when the packing fraction is small, typically of
the order of $1\%$ or less, such that one can assume 
instantaneous inelastic binary collisions with restitution coefficient $\alpha \le 1$ (the value $1$ is for elastic collisions). In experiments, usually done under gravity, it is
necessary to shake the container with accelerations much larger than
gravity in order to keep the packing fraction small everywhere ~\cite{c90,poeschel,puglio15,garzo2019granular}. The three
main categories of gas regimes are: 1) cooling granular gases, non-steady states which are initially prepared as at equilibrium, and leaving the total energy dissipate under repeated inelastic collisions~\cite{h83,BMC96,NE98}; 2) boundary driven gases, where at least one wall injects energy into the gas (e.g. vibration in experiments, thermostats in theory), reaching a non-homogeneous steady state~\cite{meerson2,lohse07,PGVP16}; 3) bulk driven granular gases, where each particle is in repeated
contact with some source of energy, for instance bouncing above a vibrating rough plate~\cite{NETP99,PLMPV98,OU98,puglisi11,puglisi12}, reaching a homogeneous steady state.

In granular gases it is customary to define a kinetic
``granular temperature''~\cite{ogawa80,kumaran98,BBDLMP05,gold08}
\begin{equation}
k_B T_g = \frac{m \langle |{\bf v}|^2 \rangle }{d},
\end{equation}
with ${\bf v}$ the velocity of each particle, $d$ the dimensionality
of space and $k_B$ is usually replaced with $1$. Such  a temperature is not expected to have a wide thermodynamic meaning, and also in statistical mechanics it has not a role equivalent to that played for molecular gases, for instance deviations from a Maxwellian are inevitable in the presence
of inelastic collisions, a
kurtosis excess (or second Sonine coefficient) is observed - larger or smaller - in many
regimes~\cite{NE98,TPNE01}.  In all gas and liquid regimes, moreover,  there  is no 
equipartition of energy among different degrees of freedom (e.g. in a mixture or under non-isotropic external forces), unless they have identical properties ~\cite{garzo1999homogeneous,MP99,wildman2002coexistence,FM02,BT02,MP02,PMP02,MP02b}.

Linear response relations have been frequently studied for granular
gases and liquids, particularly in steady
states~\cite{PBL02,DB02,BLP04,G04,SBL06,PBV07,BGM08,VPV08,VBPV09,GPSV14},
while a few studies also considered cooling
regimes~\cite{dufty2001mobility,DB02,BGM08}. In dilute homogeneously driven granular
gases, the equilibrium FDR is empirically observed, provided that the
canonical temperature is replaced with the tracer granular temperature
$T_0$ which - in general - can be different from $T_g$~\cite{PBL02,BLP04,G04,VPV08,VBPV09}.  For instance, a granular
tracer under the action of a weak perturbing force in a dilute driven
granular system satisfies the dynamical Einstein relation,
Eq.~\eqref{einst2} with $T=T_0$.  Such a result is surprising as, on
the basis of the FDR discussed above, Eq.~\eqref{LN} and of the
non-Gaussian distribution of velocities, one would expect a correction
to it. Nevertheless, in many different dilute cases, such corrections
are not observed or - in certain solvable models - can even be proven
to vanish~\cite{VBPV09}. A possible explanation to such a general
result comes through the {\em molecular chaos} which is likely to be
valid in dilute cases and which implies that a particle $1$ meets
particle $2$ only once: any collision rule, if restricted to a single
particle (that is, disregarding the fate of particle $2$) is
equivalent to an {\em elastic} collision with effective
masses~\cite{temprev}. For a massive intruder (mass much larger than
the other particles), the validity of the Einstein relation is
recovered in the context of the derivation of an effective Langevin
equation model~\cite{sarra10}

The liquid (non dilute) case is perhaps more interesting. The first experiment focusing on a Brownian-like description of a
large intruder in a granular liquid is discussed in
Ref.~\cite{DMGBLN03}. Most recent
studies, both theoretical~\cite{PBV07,VPV08,VBPV09,sarra10b,sarra12}
and experimental~\cite{GPSV14}, have shown that when the granular is a
liquid and not a gas, deviations from the equilibrium
Fluctuation-Dissipation relation are observed. In granular liquids, as a matter of fact, granular
temperature is much less useful than in gases, and cannot be replaced
by some other temperature for the purpose of an effective description.

\begin{figure}
  \centering
\includegraphics[width=10cm,clip=true]{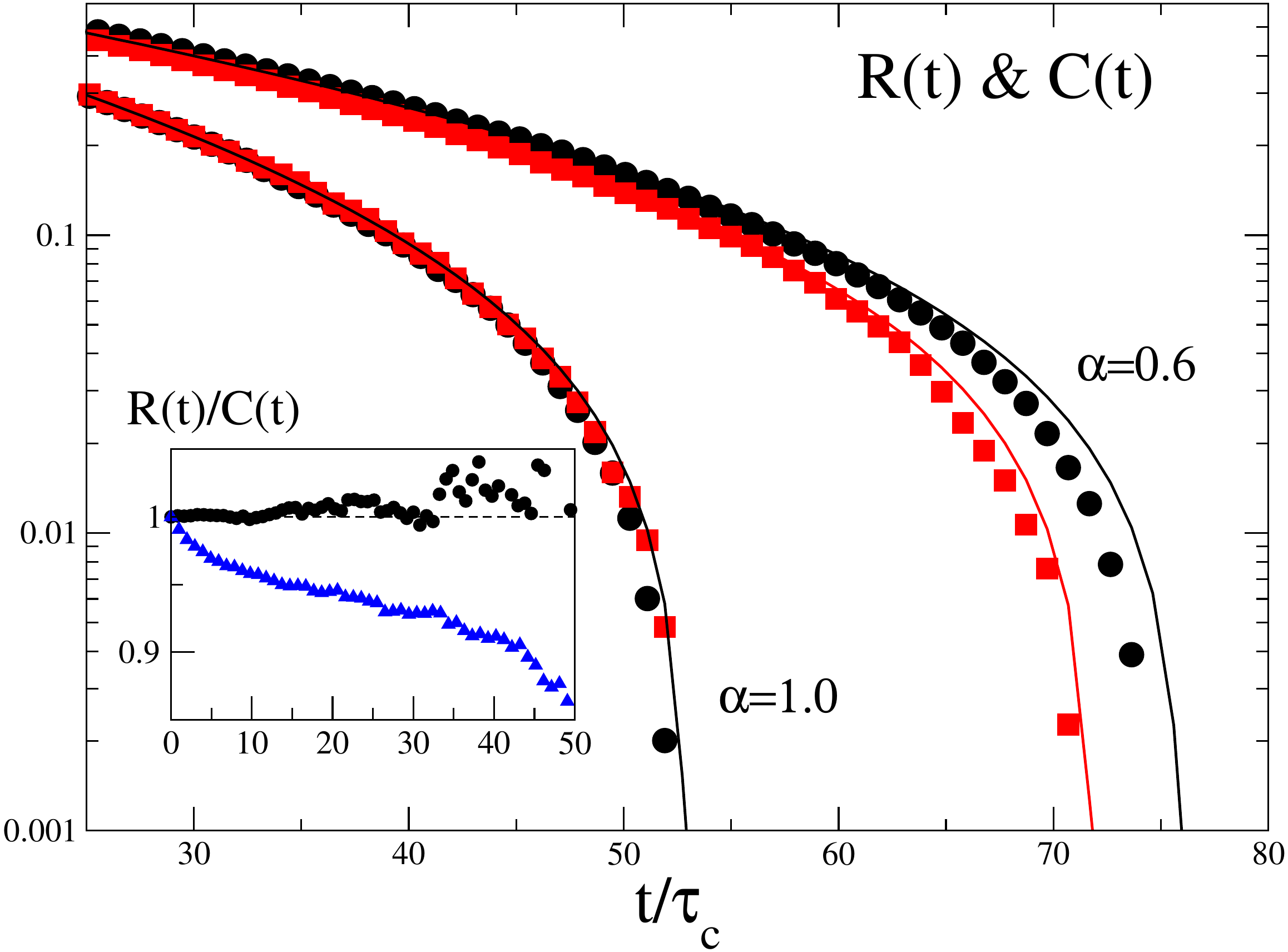}
\caption{Response function of the tracer's velocity $V$ under a perturbing force $F$, $R_{vF}(t)$ and auto-correlation
  function $C(t)=\langle V(t)V(0)\rangle/\langle V(0)V(0)\rangle$ as a
  function of time, measured in molecular dynamics simulations of a
  system composed of a massive intruder interacting with a driven
  granular fluid~\cite{sarra10b}. In the main plot an elastic case with restitution coefficient  $\alpha=1$ (where the two functions superimpose as in equilibrium FDR) and an inelastic case $\alpha<1$ (where  equilibrium FDR is violated) are shown. In the inset the ratio between the two curves is shown for the two cases (black is elastic, blue is inelastic).}
\label{fig_rc}
\end{figure}

An interesting example, in theory and in experiments is provided, again, by a massive
intruder $M \gg m$~\cite{sarra10b,sarra12}. For the purpose of
describing, in numerical simulations, the  autocorrelation of the velocity $V$ of the tracer and its linear
response, the  following model  provides a fair description for packing
fractions smaller than $40\%$: 
\begin{subequations} \label{grintr}
\begin{align} 
M \dot{V}(t)= - \Gamma [V(t)-U(t)] + \sqrt{2 \Gamma T_{tr}} \mathcal{E}_v(t)\\
M' \dot{U}(t) = -\Gamma' U(t) - \Gamma V(t) + \sqrt{2 \Gamma' T_b} \mathcal{E}_U(t),
\end{align}
\end{subequations}
where $U(t)$ is an auxiliary variable representing the memory effect due to the average velocity field of the particles surrounding the tracer, $\Gamma$ and $T_{tr}$ are the effective drag coefficient and
tracer temperature (both can be derived by kinetic theory in the
dilute limit), $\Gamma'$ and $M'$ are parameters to be determined, for
instance from the measured autocorrelation function, and $T_b$ is the
value of $T_g$ in the elastic limit (for instance the external bath
temperature~\cite{PLMPV98}). Equations~\eqref{grintr} can be mapped
into a generalised Langevin equation, Eq.~\eqref{gle}, with
exponential memory kernel. In the dilute limit (parameters such that
$U$ is negligible) the massive tracer evolves according to a simple
Langevin equation. In the elastic limit ($T_{tr} = T_b=T_g$), on the
other side, the coupling with $U$ is still important, but the
equilibrium Fluctuation-Dissipation relation is recovered. The
numerical simulations have shown that the auxiliary field $U(t)$ is a
local average of the velocities of the particles surrounding the
intruder. When the density increases numerical simulations suggest
$T_{tr} \to T_g$, likely due to a reduction of effective inelasticity
in recolliding particles. The appearance of $T_b$ is also interesting:
the “temperature” associated to the local velocity field $U$ is equal
to the bath temperature and this seems a consequence of the
conservation of momentum in collisions, implying that the average
velocity of a group of particles is not changed by collisions among
themselves and is only affected by the external bath and a (small)
number of collisions with outside particles. Summarizing,
model~\eqref{grintr} suggests that in a granular liquid - at some
level of approximation - {\em two} temperatures are relevant, one
related to the single particle scale and another one related to a
many-particle, or collective, scale. Such a conclusion is consistent
with a series of recent results about spatial velocity correlations,
typically measured as structure factors of the velocity
field~\cite{BMG09,trizac09,NEBO97,BMP02,BMP02b,NETP99,puglisi11,gradenigo11,puglisi12,PLH12,GCR13}.

\subsection{Application to biological systems and active particles}
\label{applact}

The results of the FDR have been also applied to several biological
systems, for instance in an evolution experiment in
bacteria~\cite{Sato14086} or in the prediction of heart rate
response~\cite{chen}.  Another recent application has been proposed in
the context of brain activity.  Indeed, one can wonder whether, at
some scale, the evoked activity in the brain to an external stimulus
can be somehow predicted from the observation of the spontaneous, rest
activity. In order to quantitatively address this issue, one needs an
effective model to describe the brain dynamics at the considered
scale. In the work~\cite{PhysRevResearch.2.033355}, the authors
considered the stochastic version of the Wilson-Cowan
model~\cite{benayoun2010avalanches}, describing at a coarse-grained
level the dynamics of populations of exitatory and inhibitory
neurons. In the linearized version, this model consists in two coupled
linear Langevin equations for the two populations.  The prediction of
the FDR for this model was compared to experimental
Magnetoencephalography (MEG) data for rest and evoked activity in
healthy subjects. Whereas the behavior of the temporal autocorrelation
function of the total rest activity (exitatory plus inhibitory
neurons) showed a double exponential decay characterized by two
typical times, the decay of the response function was described by a
single exponential decay, in qualitative agreement with the prediction
of the FDR. These results suggest that some information of the brain
response to external stimuli can be obtained from the observation of
its spontaneous activity.

A different field which is in large part contained in biology and
biophysics, is that of self-propelled particles, where non-equilibrium
stochastic dynamics has been employed as a main modelling tool
~\cite{marchetti2013hydrodynamics, gompper20202020}.  These systems,
known as ``active'', are usually out of equilibrium and store energy
from the environment, for instance taking advantage of chemical
reactions or mechanical agents (such as bacterial cilia and flagella),
to produce directed motion
~\cite{bechinger2016active}. 
The intrinsic non-equilibrium nature of the class of models proposed to describe active systems makes them the ideal platforms to test any version of the generalized FDR~\cite{brady,sarracino2019fluctuation}.
Since their steady-state properties are quite rich, involving unexpected spatial correlations in density, velocity and polarization fields, the use of Eq.~\eqref{LN} can be challenging.
For this reason, this method has been applied only in the limit of small activity~\cite{caprini2018linear} when the steady probability distribution is known perturbatively or using effective equilibrium-like approaches.
This allows one to derive a near-equilibrium expression for the susceptibility~\cite{fodor2016far} and approximated predictions for the transport coefficients of active particles, such as their mobility~\cite{dal2019linear}. 
In addition, the Malliavin weight sampling has been recently generalized to the more common models used to describe the active particle dynamics~\cite{szamel}.
This technique was particularly useful to explore numerically far from equilibrium regimes, calculating i) the mobility of an interacting active system at low density~\cite{dal2019linear} ii) the response function due to a shear flow~\cite{asheichyk2019response} and, finally, iii) the active effective temperature~\cite{levis2015single, nandi2018effective, cugliandolo2019effective, petrelli2020effective}. 

In this section, going beyond the approximated approaches explained so far, we apply the technique reported in Sec.~\ref{SubSec:CapriniStatement} to obtain exact expressions for the generalized FDR valid in active matter systems~\cite{caprini2020fluctuation, caprini2021generalized}.
Specifically, we focus on particle systems in the framework of dry active matter without momentum conservation.
In this context the evolution of an active particle of mass $m$ is described by a set of stochastic equations for its position, $\mathbf{x}$, and its velocity, $\mathbf{v}$, given by~\cite{mandal2019motility, caprini2021inertial}:
\begin{subequations}
\label{eq:dynamics_activecolloids_under}
\begin{align}
\dot{\mathbf{x}} &= \mathbf{v}\\
m\dot{\mathbf{v}} &= -\gamma \mathbf{v} -\nabla U + \mathbf{f}^a + \sqrt{2T\gamma} \,\boldsymbol{\eta} \,, 
\end{align}
\end{subequations}
while, in the more common overdamped version, such that $m/\gamma\ll 1$, reads:
\begin{flalign}
\label{eq:dynamics_activecolloids_over}
\gamma\dot{\mathbf{x}}  =  \mathbf{F} + \mathbf{f}^a + \sqrt{2T\gamma} \,\boldsymbol{\eta} \,. 
\end{flalign}
In both the dynamics, $\mathbf{f}^a$ is a non-gradient force, called "active force" for simplicity, that
 models at a coarse-grained level the system-dependent mechanism responsible for the active dynamics so that its complex physical or biological origin is not explicitly considered.
%
This term is chosen as a time-dependent force that provides a certain
degree of persistence to the particle trajectory in agreement with the
experimental observations of active colloids, bacteria, and other
biological microswimmers.  The most popular models to account for this
persistence in the framework of continuous stochastic processes are
the Active Brownian Particles (ABP)~\cite{buttinoni2013dynamical,
  solon2015pressure, stenhammar2014phase, farage2015effective,
  digregorio2018full, breoni2020active, caprini2020hidden} and the
Active Ornstein-Uhlenbeck particles
(AOUP)~\cite{wittmann2017effective, caprini2018active,
  maggi2017memory, dabelow2019irreversibility, berthier2019glassy,
  martin2020statistical}.  In both cases, the active force is
expressed as:
\begin{equation}
\label{eq:activeforze_def}
\mathbf{f}^a=\gamma v_0 \mathbf{n} \,,
\end{equation}
where $v_0$ is the swim velocity induced by the active force and $\mathbf{n}$ is a vector representing the particle orientation that evolves stochastically.
In the ABP model, $\mathbf{n}$ is a unit vector that evolves as
\begin{equation}
\label{eq:orientation_ABP}
\dot{\mathbf{n}}= \sqrt{2 D_r} \mathbf{n} \times \boldsymbol{\xi} \,,
\end{equation}
while in the AOUP model, $\mathbf{n}$ follows an Ornstein-Uhlenbeck process with unitary variance:
\begin{equation}
\label{eq:orientation_AOUP}
\tau\dot{\mathbf{n}}= -\mathbf{n} + \sqrt{2 \tau} \boldsymbol{\xi} \,.
\end{equation}
In both equations, $\boldsymbol{\xi}$ is a vector of $\delta$-correlated white noises with zero average.
The coefficient $D_r$ is the rotational diffusion coefficient while $\tau$ is simply named persistence time since it coincides with the autocorrelation time of the active force.
The models reproduce consistent results by choosing $(d-1)D_r=1/\tau$ where $d>1$ is the dimension of the system~\cite{farage2015effective}.

In general, the active force pushes the system out of equilibrium, producing entropy with  a rate that grows with $\tau$ ~\cite{caprini2019entropy, mandal2017entropy, shankar2018hidden, chaki2019effects, dabelow2020irreversible}. 
Applying Eq.~\eqref{eq:R_capriniadditive} to the dynamics~\eqref{eq:dynamics_activecolloids_under}, the elements of the response matrix after perturbing the $x$ component of the velocity, read~\cite{caprini2021generalized}:
\begin{subequations}
\label{eq:active_underdamped_x}
\begin{align}
\label{eq:FDRdiag_activeunder}
&\mathcal{R}_{v, v}(t) = \frac{m}{T } \left\langle v(t) v(0) \right\rangle + \frac{m}{2 T \gamma} \left(\left\langle v(t) \nabla_x U(0) \right\rangle + \left\langle \nabla_x U(t) v(0) \right\rangle  - \left\langle v(t) \mathrm{f}^a(0) \right\rangle - \left\langle \mathrm{f}^a(t) v(0) \right\rangle  \right) \\
\label{eq:FDRnondiag_activeunder}
&\mathcal{R}_{x, v} (t) = \frac{m}{ 2T }  \langle x(t) v(0)  \rangle +  \frac{m}{2 T \gamma } \langle x(t) \nabla_x U(0) \rangle -  \frac{m}{2 T \gamma } \langle x(t) \mathrm{f}^a(0) \rangle  - \frac{m^2}{2T\gamma}\langle v(t) v(0) \rangle \,,
\end{align}
\end{subequations}
where we have suppressed the spatial indices for simplicity.
Equation~\eqref{eq:FDRdiag_activeunder} is determined by the generalized retarded kinetic energy and the time-symmetric retarded power injected by the gradient force and the active force.
In Eq.~\eqref{eq:FDRnondiag_activeunder}, we can identify the retarded mechanical pressure (second term), the so-called retarded swim/active pressure (third term) and, finally, the retarded/anticipated kinetic energy (fourth term).
Applying Eq.~\eqref{eq:R_capriniadditive} to the dynamics~\eqref{eq:dynamics_activecolloids_over}, the response after perturbing the coordinate $x$ of the particle position reads~\cite{caprini2021generalized}:
\begin{equation}
\label{eq:Resp_x_overdamped_active}
\mathcal{R}_{x, x}(t) =   \frac{1}{2 T } \left(\left\langle x(t) \nabla_{x} U(0) \right\rangle + \left\langle \nabla_{x}U(t) x(0) \right\rangle  \right) - \frac{1}{2T } \left(\left\langle x(t) \mathrm{f}^a(0) \right\rangle + \left\langle \mathrm{f}^a(t) x(0) \right\rangle  \right) \,.
\end{equation}
In the overdamped case, the response is determined by the sum of the time-symmetric part of the retarded/anticipated mechanical and swim pressures. 
In overdamped systems with $T=0$, the above formulation of the FDR cannot be directly applied, because the dynamics is not of the Langevin form. 
In this athermal case, another version of the generalized FDR can be derived using a modified path-integral method developed in~\cite{caprini2020fluctuation} in the case of AOUP (FDR for athermal ABP are still unknown), obtaining:
\begin{flalign}
\label{eq:second_result}
D_a\gamma\mathcal{R}_{x, x}(t) =&\frac{1}{2}\left[ \left\langle x(t) \nabla_{x} U(0) \right\rangle +\left\langle \nabla_{x} U(t) x(0) \right\rangle  \right] \nonumber\\
&\qquad+\frac{\tau^2}{2}\sum_{\alpha}\left[ \left\langle v_{\alpha}(t) \nabla_{\alpha}\nabla_{x} U(t) v_x(0) \right\rangle +\left\langle v_x(t) \nabla_{x}\nabla_{\alpha} U(0) v_{\alpha}(0) \right\rangle  \right] \,,
\end{flalign}
where we have introduced the particle velocity $v_{\alpha}=\dot{\alpha}$, with $\alpha=x,y$.
According to our notation, repeated indices are summed, $U(s)=U(\mathbf{x}(s))$. 
The first line of Eq.~\eqref{eq:second_result} coincides with the equilibrium FDR holding for passive particles where the detailed balance holds.
The second line contains two additional terms, involving the particle velocity and the second derivative of the potential, that disappears in the equilibrium limit $\tau\to 0$. 
At variance with the equilibrium scenario, in athermal active systems, the generalized FDR is not only determined by a time correlation involving the position but is  affected by the correlations between the other variables, such as the velocity.  

\begin{figure}[!t]
\centering
\includegraphics[width=0.95\linewidth,keepaspectratio]{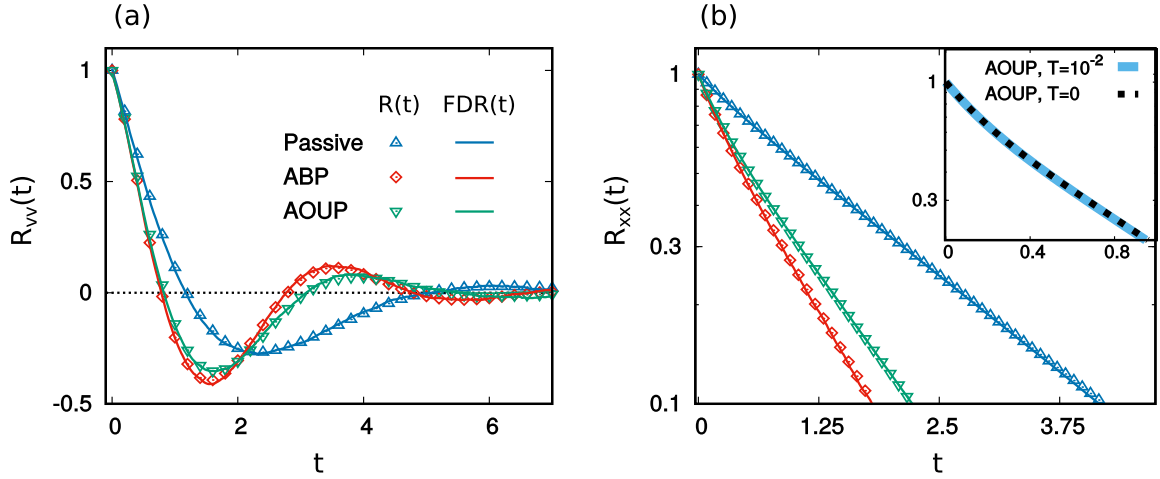}
\caption{\label{fig:figure1}
Comparison between response and FDR for a two-dimensional particle confined in a quartic potential, $U(\mathbf{x})=k|\mathbf{x}|^4$. 
Panel (a): $R_{vv}(t)$ (colored points) calculated perturbing the velocity of the underdamped dynamics, Eq.~\eqref{eq:dynamics_activecolloids_under}. Panel (b): $R_{xx}(t)$ (colored points) calculated perturbing the position of the overdamped dynamics, Eq.~\eqref{eq:dynamics_activecolloids_over}. The responses are shown for passive, ABP and AOUP as explained in the legend which is shared by both panels. Solid color lines plot the FDR, obtained using Eq.~\eqref{eq:FDRdiag_activeunder} and~\eqref{eq:Resp_x_overdamped_active}, for panels (a) and (b), respectively. 
The inset of panel (b) shows a comparison between Eq.~\eqref{eq:Resp_x_overdamped_active} (calculated at $T=10^{-2}$) and Eq.~\eqref{eq:second_result} (holding for $T=0$).
The other parameters of the simulations are $k=3$, $\gamma=1$, $T=10^{-1}$, $v_0=1$, and $\tau=1$.
}
\end{figure}

To validate the generalized FDR in the case of active particles, we consider both AOUP and ABP dynamics confining the particle through a non-linear force due to an external potential. 
To go beyond the harmonic case that can be solved analytically~\cite{caprini2020fluctuation}, we chose a quartic potential, $U(\mathbf{x})=k|\mathbf{x}|^4/4$, where the constant $k$ determines the strength of $U$.
In Fig.~\ref{fig:figure1}, we show the diagonal elements of the response matrix numerically obtained by their definitions (i.e. perturbing the dynamics) and the FDR numerically calculated from the unperturbed system.
In particular, in panel~(a), we show the results in the underdamped case, reporting the profile of $R_{vv}(t)$ and the FDR calculated from Eq.~\eqref{eq:FDRdiag_activeunder}, while, in panel~(b), the analogue study is reported for the overdamped dynamics, comparing $R_{x,x}(t)$ and the FDR, Eq.~\eqref{eq:Resp_x_overdamped_active}. 
In both cases, the FDRs exactly match with the direct study of the response confirming the exactness of our theoretical results.
Finally, in the inset of panel~(b), we compare Eq.~\eqref{eq:Resp_x_overdamped_active} in the limit of small temperature, $T$, and the athermal relation, Eq.~\eqref{eq:second_result}.
We reveal that the former converges onto the latter for $T\to0$.

\section{Conclusions}
\label{concl}

We have reviewed two significant approaches to the problem of linear
response in general systems, when the constraint of thermodynamic
equilibrium for the unperturbed state is removed. We have also
sketched some of the interesting recent applications of such
approaches. We cannot avoid to stress, again, the evident fact that -
given the system, the observable of interest and the applied
perturbation - the linear response function is unique and therefore
the two approaches lead to the same result, and in fact an analytical
connection can be demonstrated~\cite{bcvunp}. The difference between
the two schemes relies on the required information: in one case,
formula~\eqref{LN}, one needs some knowledge about the probability
distribution at initial time (e.g. the steady-state one) for the
relevant degrees of freedom; in the other case,
formulas~\eqref{eq:Novikovequation},~\eqref{eq:R_capriniadditive}
and~\eqref{discreto}, one needs knowledge about the system's dynamical
model (e.g. noise distributions, forces involved, transition rates,
etc.). It is not always evident when one approach is more useful than
the other. In lucky cases, where both the dynamical model and its
probability distribution are known, the two formulas can express
different information and one can be more useful than the other (for
instance correlations with state variables can be more transparent
than correlations with noises or time-derivative of state variables).

 In experimental situations, where the underlying model is not known,
 an empirical approach to retrieve the main features of the
 probability distribution of the relevant degrees of freedom can be
 simpler than retrieving infomation about forces and noises in the
 system, suggesting the first approach as the more useful.  If a
 dynamical model is known for the relevant degrees of freedom, while
 the generated probability distribution is unknown, then the second
 approach should be more direct. However it is clear that, even when a
 dynamical model is fully available, the first approach may have some
 advantage: for instance, in a system with many particles and a
 massive tracer whose response is investigated, the knowledge of the
 dynamics of all the particles can be too detailed and result, when
 inserted in the second approach, in quite a complicate formula, or
 even not very informative and/or transparent ones; an empirical study
 of the probability distribution of the relevant degrees of freedom
 (e.g. those of the tracer and some coarse-grained observable for the
 surrounding fluid) can provide, sometimes, an approximate but more informative route
 through the first approach (see for instance the example discussed in
 Section~\ref{sub51}).

We also recall that an FDR does not give an explicit prediction for
the response, but only an expression of it in terms of unperturbed
correlations. Once an FDR is known, the problem of obtaining
(empirically or analytically) the required correlations remains. An
FDR however can have already a theoretical meaning, even without the
explicit knowledge of the time-dependence of the involved unperturbed
correlations, i.e. it is already significant to know {\em which}
correlations are involved, as well illustrated by the application
described in subsection~\ref{bald} for the problem of causation and
also in the closure problem in the Kraichnan's approach to
turbulence~\cite{Kr59}.


\begin{acknowledgement}
The Authors acknowledge partial financial support from MIUR through the PRIN 2017 grant number 201798CZLJ. AP also acknowledges financial support from Regione Lazio through the Grant “Progetti Gruppi di Ricerca” N. 85-2017-15257.
\end{acknowledgement}

\bibliographystyle{spphys}
\bibliography{biblio,Respbib}

\end{document}